\documentstyle[12pt]{article}
\def\le{\langle}
\def\re{\rangle}

\def\1{\mbox{I\hspace{-.15em}1}}

\def\b{\begin{equation}}
\def\e{\end{equation}}
\def\bee{\begin{enumerate}}
\def\eee{\end{enumerate}}

\title{ Negative Norm States in de Sitter Space and \\QFT without Renormalization Procedure}
\author{ Mohammad Vahid Takook \thanks{e-mail: takook@ccr.jussieu.fr; takook@razi.ac.ir}}
\date{\today}

\begin{document}
\maketitle {\it \centerline{Department of Physics, Razi
University, Kermanshah, IRAN}}

\begin{abstract}

In recent papers \cite{dere, gareta1}, it has been shown that the
presence of negative norm states or negative frequency solutions
are indispensable for a fully covariant quantization of the
minimally coupled scalar field in de Sitter space. Their presence,
while leaving unchanged the physical content of the theory, offers
the advantage of eliminating any ultraviolet divergence in the
vacuum energy \cite{gareta1} and infrared divergence in the two
point function \cite{ta3}. We attempt here to extend this method
to the interacting quantum field in Minkowski space-time. As an
illustration of the procedure, we consider the $\lambda\phi^4$
theory in Minkowski space-time. The mathematical consequences of
this method is the disappearance of the ultraviolet divergence to
the one-loop approximation. This means, the effect of these
auxiliary negative norm states is to allow an automatic
renormalization of the theory in this approximation.

\end{abstract}

\vspace{0.5cm} {\it Proposed PACS numbers}: 04.62.+v, 03.70+k,
11.10.Cd, 98.80.H \vspace{0.5cm}

\section{Introduction}

Antoniadis, Iliopoulos and Tomaras \cite{anilto} have shown that
the pathological large-distance behavior (infrared divergence) of
the graviton propagator on a de Sitter background does not
manifest itself in the  quadratic part of the effective action  in
the one-loop approximation. This means that the pathological
behavior of the graviton propagator may be gauge dependent and so
should not appear in an effective way as a physical quantity. The
linear gravity (the traceless rank-2 ``massless'' tensor field) on
de Sitter space is indeed built up from copies of the minimally
coupled scalar field \cite{ta1,gareta2}. It has been shown
\cite{dere, gareta1} that one can construct a covariant
quantization of the ``massless'' minimally coupled scalar field in
de Sitter space-time, which is causal and free of any infrared
divergence. The essential point of that paper is the unavoidable
presence of the negative norm states. Although they do not
propagate in the physical space, they play a renormalizing role.
In the forthcoming paper \cite{gareta2}, we shall show that this
is also true for linear gravity (the traceless rank-2 ``massless''
tensor field). These questions have recently been studied by
several authors (for minimally coupled scalar field see
\cite{ta1,vera,gareta1,ta3} and for linear gravity see
\cite{ta1,ta2,hahetu,hiko}).

The auxiliary states (the negative norm states) appear to be
necessary for obtaining a fully covariant quantization of the free
minimally coupled scalar field in de Sitter space-time, which is
free of any infrared divergence. It has been shown that these
auxiliary states automatically renormalize the infrared divergence
in the two-point function \cite{ta3} and removes the ultraviolet
divergence in the stress tensor \cite{gareta1}. The crucial point
about the minimally coupled scalar field lies in the fact that
there is no de Sitter invariant decomposition $${\cal H}={\cal
H}_+\oplus{\cal H}_-,$$ where ${\cal H}_+$ and ${\cal H}_-$ are
Hilbert  and anti-Hilbert spaces respectively. For this reason our
states contain negative frequency solutions and consequently the
use of a Krein space ({\it i.e.} Hilbert $\oplus$ anti-Hilbert
space) is necessitated. For the scalar massive field where such a
decomposition exists as a de Sitter invariant, ${\cal H}_+$ as the
usual physical state space $({\cal H}_{-}={\cal H}_{+}^{*})$
suffices \cite{gareta1}. It has been also shown that if this
method is applied to the free ``massive'' scalar field in de
Sitter space, automatically covariant renormalization of the
vacuum energy divergence is obtained \cite{gareta1}. We would like
to generalize this method (adding the negative frequency
solutions) to the interacting quantum scalar field in Minkowski
space-time. These auxiliary states once again, automatically
renormalize  the problem to the one-loop approximation. In other
words, introducing negative frequency solutions plays the key role
in the renormalization procedure.

\section{de Sitter scalar field}

 Let us briefly describe our quantization of the minimally coupled massless scalar field.
 It is defined by $$\Box_H \phi(x)=0,$$
 where $\Box_H$ is the Laplace-Beltrami operator on de Sitter space.
 As proved by Allen \cite{al}, the covariant canonical quantization
 procedure with positive norm states fails in this case. The
 Allen's result can be reformulated in the following way: the Hilbert space generated by
 a complete set of modes (named here the positive modes, including the zero mode) is
 not de Sitter invariant,
 $${\cal H}=\{\sum_{k \geq 0}\alpha_k\phi_k;\;
 \sum_{k \geq 0}|\alpha_k|^2<\infty\}.$$
  This means that it is not closed under the action of the de~Sitter group.
 Nevertheless, one can obtain a fully covariant quantum field by adopting
 a new construction \cite{dere,gareta1}. In order to obtain a fully covariant
 quantum field, we add all the conjugate modes to the previous ones.
 Consequently, we have to deal with an orthogonal sum of a positive and
 negative inner product space, which is closed under an indecomposable
 representation of the de~Sitter group. The negative values of the inner
 product are precisely produced by the conjugate modes:
 $\le\phi_k^*,\phi_k^*\re=-1$, $k\geq 0$. We do insist on the fact that the
 space of solution should contain the unphysical states with negative norm.
  Now, the decomposition of the field operator into positive and negative norm parts reads
   \b \phi(x)=\frac{1}{\sqrt{2}}\left[ \phi_p(x)+\phi_n(x)\right],\e
 where
 \b \phi_p(x)=\sum_{k\geq 0} a_{k}\phi_{k}(x)+H.C.,\;\;
  \phi_n(x)=\sum_{k \geq 0} b_{k}\phi^*(x)+H.C..\e
 The positive mode $\phi_p(x)$ is the scalar field as was used by Allen.
 The crucial departure from the standard QFT based on CCR lies in the
 following requirement on commutation relations:
 \b    a_{k}|0>=0,\;\;[a_{k},a^{\dag} _{k'}]= \delta_{kk'},\;\;
   b_{k}|0>=0,\;\;[b_{k},b^{\dag} _{k'}]= -\delta_{kk'}.\e
 A direct consequence of these formulas is the positivity of the energy  {\it i.e.}
$$\le\vec k|T_{00}|\vec k\re\geq0,$$ for any physical state $|\vec
k\re$ (those built from repeated action of the $a^{\dag} _{k}$'s
on the vacuum). This quantity vanishes if and only if $|\vec
k\re=|0\re$. Therefore the ``normal ordering'' procedure for
eliminating the ultraviolet divergence in the vacuum energy, which
appears in the usual QFT is not needed \cite{gareta1}. Another
consequence of this formula is a covariant two-point function,
which is free of any infrared divergence \cite{ta3}.

This result is the same as that of de Vega et al. \cite{vera}
where flat coordinate modes solutions were employed. For
calculating the Schwinger commutator function, they have not used
the two-point function since in their construction it would result
in appearance of a divergence. They calculated the commutator
function directly, which resulted in disappearance of the infrared
divergence due to the sign of the divergence term. In our
alternative method, the Schwinger commutator function was
calculated from the finite and covariant two point function
\cite{ta3}. It has been also shown that the Schwinger commutator
functions in both methods are one and the same.

\section{ Minkowskian free quantum scalar field}

Let us first recall elementary facts about Minkowskian QFT. A
classical scalar field $\phi(x)$, which is defined in the
$4$-dimensional Minkowski space-time, satisfies the field equation
\b (\Box+m^2)\phi(x)=0=(\eta^{\mu\nu}\partial_\mu
\partial_\nu+m^2)\phi(x),\;\;
\eta^{\mu\nu}=\mbox{diag}(1,-1,-1,-1).\e Inner or {\it
Klein-Gordon} product and related norm are defined by \cite{bida}
\b
(\phi_1,\phi_2)=-i\int_{t=\mbox{cons.}}\phi_1(x)\stackrel{\leftrightarrow}
{\partial}_t\phi_2^*(x)d^3x.\e  Two sets of solutions of $(4)$ are
given by: \b u_p(k,x)=\frac{e^{i\vec k.\vec
x-iwt}}{\sqrt{(2\pi)^32w}}
=\frac{e^{-ik.x}}{\sqrt{(2\pi)^32w}},\;\;u_n(k,x)=\frac{e^{-i\vec
k.\vec x+iwt}}{\sqrt{(2\pi)^32w}}
=\frac{e^{ik.x}}{\sqrt{(2\pi)^32w}},\e where $ w(\vec k)=k^0=(\vec
k .\vec k+m^2)^{\frac{1}{2}} \geq 0$.  These $u(k,x)$  modes are
orthogonal and normalized in the sense of $(5)$: $$
(u_p(k,x),u_p(k',x))=\delta^3(\vec k-\vec k'),$$ $$
(u_n(k,x),u_n(k',x))=-\delta^3(\vec k-\vec k'),$$ \b
(u_p(k,x),u_n(k',x))=0.\e $u_p$ modes are positive norm states and
the $u_n$'s are negative norm states. The general classical field
solution is $$ \phi(x)=\int d^3\vec k [a(\vec k)u_p(k,x)+b(\vec
k)u_n(k,x)],$$ where $a(\vec k)$ and $b(\vec k)$ are two
independent coefficients. The usual quantization of this field is
based on the positive norm states only. In the Minkowskian case
this choice leads to a covariant quantization (covariant under the
proper orthochronous Poincar\'e group). However, it is well known
that an ultraviolet divergence appears in the vacuum energy. This
divergence is eliminated with the aid of a ``normal ordering''
operation.

In the above, we have seen that, in the case of the minimally
coupled scalar field in de Sitter space, one cannot construct a
covariant quantization of this field with only positive norm
states (this fact was proved by Allen in  \cite{al}). Also there
appears an infrared divergence in the two-point function built
from the positive norm states. For obtaining a covariant
quantization and eliminating the infrared divergence the two sets
of solutions (positive and negative norms states) are necessary
\cite{gareta1}. It has been also shown that the commutator
function, which is calculated by these two different methods, is
the same \cite{ta3}. Therefore there exists another possibility
for defining the field operator, which satisfies the same
commutation relation (locality condition). In contrast to the
usual quantization, the field operator acts on the Krein space
(positive and negative norms states). Let us show now that, if we
use this new method of quantization for the free scalar field in
Minkowski space, the ultraviolet divergence in the vacuum energy
disappears and there is no need for use the ``normal ordering''
operation. In Krein QFT the quantum field is defined as follows \b
\phi(x)=\frac{1}{\sqrt 2}[\phi_p(x)+\phi_n(x)],\e where $$
\phi_p(x)=\int d^3\vec k [a(\vec k)u_p(k,x)+a^{\dag}(\vec
k)u_p^*(k,x)],$$ $$ \phi_n(x)=\int d^3\vec k [b(\vec
k)u_n(k,x)+b^{\dag}(\vec k)u_n^*(k,x)],$$ where $a(\vec k)$ and
$b(\vec k)$ are two independent operators. The positive mode
$\phi_p$ is the scalar field as was used in the usual QFT.
Creation and annihilation operators are constrained to obey the
following commutation rules \b [a(\vec k),a(\vec
k')]=0,\;\;[a^{\dag}(\vec k), a^{\dag}(\vec k')]=0,\;\;, [a(\vec
k),a^{\dag}(\vec k')]=\delta(\vec k-\vec k') ,\e \b [b(\vec
k),b(\vec k')]=0,\;\;[b^{\dag}(\vec k), b^{\dag}(\vec k')]=0,\;\;,
[b(\vec k),b^{\dag}(\vec k')]=-\delta(\vec k-\vec k') ,\e
 \b [a(\vec k),b(\vec k')]=0,\;\;[a^{\dag}(\vec k), b^{\dag}(\vec k')]=0,\;\;, [a(\vec
k),b^{\dag}(\vec k')]=0,\;\;[a^{\dag}(\vec k),b(\vec k')]=0 .\e
The vacuum state $\mid 0>$ is then defined by \b a^{\dag}(\vec
k)\mid 0>= \mid 1_{\vec k}>;\;\;a(\vec k)\mid 0>=0, \forall \vec
k,\e \b b^{\dag}(\vec k)\mid 0>=  \mid \bar1_{\vec k}>;\;\;b(\vec
k)\mid 0>=0, \forall \vec k,\e \b  b(\vec k)\mid 1_{\vec k}
>=0;\;\; a(\vec k)\mid \bar1_{\vec k} >=0, \forall \vec k,\e
where $\mid 1_{\vec k}> $ is called a one particle state and $\mid
\bar1_{\vec k}>$ is called a one ``unparticle state''. These
commutation relations, together with the normalization of the
vacuum $$ <0\mid 0>=1,$$ lead to positive (resp. negative) norms
on the physical (resp. unphysical) sector: \b < 1_{\vec k'} \mid
1_{\vec k}>=\delta(\vec k-\vec k'),\;\;\; < \bar 1_{\vec k'} \mid
\bar 1_{\vec k}>=-\delta(\vec k-\vec k').\e If we calculate the
energy operator in terms of these Fourier modes, we have \b H
=\int d^3\vec k k^0  [a^{\dag}(\vec k)a(\vec k)+b^{\dag}(\vec
k)b(\vec k)+a^{\dag}(\vec k)b^{\dag}(\vec k)+a(\vec k)b(\vec
k)].\e This energy for the vacuum state is zero and it is not
needed to use the ``normal ordering'' operation. It is also
positive for any particles state or physical state $\mid N_{\vec
k}>$ (those built from repeated action of the $a^{\dag}(\vec k)$'s
on the vacuum)$$ < N_{\vec k'}\mid  H \mid N_{\vec k'}>=\int  <
N_{\vec k'}\mid a^{\dag}(\vec k)a(\vec k) \mid N_{\vec k'}> k^0
d^3\vec k \geq 0,\;\;\;\mid 0_{\vec k'}>\equiv \mid 0>.$$

We shall attempt to generalize this method to the interacting
quantum field in the next section. At this stage we consider
various Green's functions fundamental to the interacting case.
Within the framework of our approach, the two-point function is
the imaginary part the usual Wightman two-point function, which is
built from the positive norm states \b {\cal W}(x,x')=<0\mid
\phi(x)\phi(x') \mid 0>=\frac{1}{ 2}[{\cal W}_p(x,x')+{\cal
W}_n(x,x')]=i \Im {\cal W}_p(x,x') ,\e where ${\cal W}_n=-{\cal
W}_p^*$. The commutator and anticommutator of the field are
defined respectively by \b iG(x,x')=<0\mid [\phi(x),\phi(x') ]\mid
0>=2i \Im {\cal W}(x,x')=2 i\Im {\cal W}_p(x,x')=iG_p(x,x') ,\e \b
G^1(x,x')=<0\mid \{\phi(x),\phi(x')\} \mid 0>=0.\e Retarded and
advanced Green's functions are defined respectively by \b
G^{ret}(x,x')=-\theta (t-t')G(x,x') =G^{ret}_p(x,x'),\e \b
G^{adv}(x,x')=\theta (t'-t)G(x,x')= G^{adv}_p(x,x') .\e The
Schwinger commutator function, retarded and advanced Green's
functions is the same in the two formalism. The ``Feynman''
propagator or the Time-ordered product propagator is defined \b
iG_T(x,x')=<0\mid T\phi(x)\phi(x') \mid 0>=\theta (t-t'){\cal
W}(x,x')+\theta (t'-t){\cal W}(x',x).\e In this case we obtain \b
G_T(x,x')=\frac{1}{2}[G_F^p(x,x')+(G_F^p(x,x'))^*]=\Re
G_F^p(x,x').\e where the positive norm state is \b
G_F^p(x,x')=\int \frac{d^4 k}{(2\pi)^4}e^{-ik.(x-x') }\tilde
G^p(k)=\int \frac{d^4
k}{(2\pi)^4}\frac{e^{-ik.(x-x')}}{k^2-m^2+i\epsilon} .\e Using the
Bessel functions it is also written in the following form \b
G_T(x,x')=\Re G_F^p(x,x')=\frac{m^2}{8\pi}\theta(\sigma)\frac{J_1
(\sqrt{2m^2\sigma})}{\sqrt{2m^2 \sigma}}-\frac{1}{8\pi}\delta
(\sigma),\e where \cite{bida} $$ G_F^p(x,x')=-\frac{1}{8\pi}\delta
(\sigma)+\frac{m^2}{8\pi}\theta(\sigma)\frac{J_1
(\sqrt{2m^2\sigma})-iN_1 (\sqrt{2m^2\sigma})}{\sqrt{2m^2
\sigma}}$$ $$ -\frac{im^2}{4\pi^2}\theta(-\sigma)\frac{K_1
(\sqrt{-2m^2\sigma})}{\sqrt{-2m^2
\sigma}},\;\;\;\sigma=\frac{1}{2}(x-x')^2.$$ The ultraviolet
singularity appears in the imaginary part of the Feynman
propagator $G_F^p(x,x)$ (equation $(9.52)$ in \cite{bida}),
$$G_F^p(x,x)\approx \lim_{n\longrightarrow
4}\frac{-2i}{(4\pi)^2}\frac{m^2}{(n-4)}+G_F^{\mbox{finite}}(x,x),$$
where $G_F^{\mbox{finite}}(x,x)$ is finite as $n\longrightarrow 4
$. Then the Green's function $G_T$ is convergence in the
ultraviolet limit \b G_T(x,x)=\Re G_F^p(x,x)=\Re
G_F^{\mbox{finite}}(x,x).\e This Green's function for every
time-like separated pair $(x,x')$ is \b \lim_{x\rightarrow
x'}G_T(x,x')= \frac{m^2}{16\pi},\e and for space-like separated
pair $(x,x')$ is zero $$ \lim_{x\rightarrow x'}G_T(x,x')=0.$$ In
the momentum space for this propagator we have \cite{itzu} \b
\tilde G(k)=\frac{1}{2}[\tilde G^p(k)+\tilde
G^p(k)^*]=\frac{1}{2}\left[\frac{1}{k^2-m^2+i\epsilon}+
\frac{1}{k^2-m^2-i\epsilon}\right]=PP \frac{1}{k^2-m^2},\e where
$PP$ is the principal part symbol.

\section{The interaction QFT}

In the interaction case the S matrix elements, which describes the
scattering of the i states into the f states ($S_{fi}=<\mbox{out,
f}|\mbox{in, i}>$) are the most important quantities to be
calculated. The S matrix elements can be written in terms of the
time order product of the two free field operator $(22)$ by
applying the reduction formulas, Wick's theorem and time evolution
operator \cite{itzu}. As this two point function is convergence in
the ultraviolet and infrared limit $(26)$, this method may be
renormalized automatically. Here the $\lambda \phi^4$ interaction
field in Minkowski space is studied to the one-loop approximation.

The tree order S-matrix elements do not change when it is applied
to the Krein QFT since the unphysical state disappear in the
external line due to the conditions $(11)$ and $(14)$  and the
internal propagator in the two cases are the same $(28)$.

In the one-loop approximation case, two primitive divergent
integrals appear, which can be written in the following form  \b
-i \Sigma^p=\frac{\lambda}{2}\int \frac{d^4k}{(2\pi)^4}
\frac{1}{k^2-m^2+i\epsilon}=\frac{\lambda}{2}\lim_{x\rightarrow
x'}G_F^p(x,x';m^2),\e
 \b \Gamma^4(s,t,u)=-i\lambda+\Gamma(s)+\Gamma(t)+\Gamma(u),\e
where $s,t$ and $u$ are the Mandelstam variable and \b
\Gamma(s)=\frac{\lambda^2}{2}\int_0^1 dl \int
\frac{d^4k}{(2\pi)^4} \frac{1}{[k^2-m^2+s l(1-l)+i\epsilon]^2}. \e
If we define $M^2=m^2-sl(1-l)$, this integral can be written in
terms of the first one $$ \Gamma(s)=\frac{\lambda^2}{2} \int_0^1
dl\frac{\partial}{\partial M^2}\int \frac{d^4k}{(2\pi)^4}
\frac{1}{k^2-M^2+i\epsilon}$$ $$=\frac{\lambda^2}{2} \int_0^1
dl\frac{\partial}{\partial M^2}\lim_{x\rightarrow
x'}G_F^p(x,x';M^2),\;\;M^2\neq 0.$$ By using the equation $(26)$
we see simply that if the Green function $G_F^p$ is replaced by
the Green function $G_T$ the integrals is convergent.

Now the ``self energy'' graph is explicitly considered. By using
the above discussion (equations $(26)$ and $(27)$) the self energy
terms of the two point function in the new approach is given by \b
-i\Sigma =\frac{\lambda}{2}\lim_{x\rightarrow
x'}G_T(x,x';m^2)=\frac{\lambda  m^2}{32\pi}, \;\; m^2\neq 0.\e The
positive norm states and the full propagator in the one loop
correction are given respectively by \b
i\Delta(k)=\frac{i}{k^2-m^2-\frac{\lambda}{32\pi}m^2+i\epsilon},\;\;\;
PP \frac{i}{k^2-m^2-\frac{\lambda}{32\pi}m^2}.\e Then in the one
loop correction the mass is replaced by \b
m^2(\lambda)=m^2(0)[1+\frac{\lambda }{32\pi}+......],\e where
$m(0)=m$. Due to the interaction, the effective mass of the
particle $m(\lambda)$, which determines its response to an
externally applied force, is certainly different from the mass of
the particle without interaction $m(0)$. In this case, $m(0)$ and
$m(\lambda)$ are both  measurable.

Finally we calculate explicitly the transition amplitude of the
state $|q_1,q_2;\mbox{ in}>$  to the state $|p_1,p_2; \mbox{
out}>$ for s-channel contribution in the one-loop approximation.
It is given by \cite{itzu} $${\cal T}\equiv<p_1,p_2; \mbox{
out}|q_1,q_2;\mbox{ in}>_s=\int d^4y_1d^4y_2d^4x_1d^4x_2
e^{ip_1.y_1+ip_2.y_2-iq_1.x_1-iq_2.x_2}$$ $$
(\Box_{y_1}+m^2)(\Box_{y_2}+m^2)(\Box_{x_1}+m^2)(\Box_{x_2}+m^2)
\frac{(-i\lambda)^2}{2!}\int d^4z_1d^4z_2 [iG_T(z_1-z_2)]^2$$ $$
G_T(y_1-z_2)G_T(y_2-z_1)G_T(x_1-z_2)G_T(x_2-z_2),$$ where the
Feynman Green function $G_F^p$ is replaced by the Time-order
product Green function $G_T$. We obtain $$ {\cal
T}=\frac{\lambda^2}{2}\int d^4z_1d^4z_2
e^{i(p_1+p_2).z_1-i(q_1+q_2).z_2}[G_T(z_1-z_2)]^2$$ \b
=\frac{\lambda^2}{2}(2\pi)^4 \delta^4(p_1+p_2-q_1-q_2)\int d^4z
e^{i(p_1+p_2).z}\left( \frac{m^2}{8\pi}\theta(z^2)\frac{J_1
(\sqrt{m^2z^2})}{\sqrt{m^2 z^2}}-\frac{1}{4\pi}\delta
(z^2)\right)^2,\e where $2\sigma=(z_1-z_2)^2=z^2$. The integral
for the space-like separated pair $(z_1,z_2)$ is zero. That means
the interaction between the intermediate states do not exist for a
space-like separated pair. Then the causality or locality
principle is preserved for the intermediate states. By using the
equation $(26)$ we obtained that the integral for the light-like
separated pair $(z_1,z_2)$ is also zero. That means an internal
``particle'', which propagate in the intermediate states, is
``massive'' and it can not propagate on the light-cone.

The integral for the time-like separated pair $(z_1,z_2)$ is
finite \b \int d^4z e^{i(p_1+p_2).z}\left(\theta(z^2)\frac{J_1
(\sqrt{m^2z^2})}{\sqrt{m^2 z^2}}\right)^2=\mbox{finite}.\e
Therefore the transition amplitude is finite in the one-loop
approximation.

\section{Conclusion}

We recall that the negative frequency solutions of the field
equation are needed for quantizing in a correct way the minimally
coupled scalar field in de Sitter space. Contrary to the Minkowski
space, the elimination of de Sitter negative norm in the minimally
coupled states breaks the de Sitter invariance. Then for restoring
the de Sitter invariance, one needs to take into account the
negative norm states {\it i.e.} the Krein space  quantization. It
provides a natural tool for renormalization technique
\cite{gareta1}. In this paper the $\lambda \phi^4$ theory in
Minkowski space-time has been studied to the one-loop
approximation in the Krein space quantization. It is found that
the theory is automatically renormalized in this approximation.
The main questions, which naturally arise, are: {\it dose the
natural renormalizeability preserve in the higher-loop expansion?}
{\it does this construction affect the physical world ?} \vskip
0.5 cm

\noindent {\bf{Acknowlegements}}: The author would like to thank
S. Rouhani for very useful discussions and  S. Teymourpoor and M.
Oveisy for their interest in this work.

\end{document}